# A New QF1 Magnet for ATF3

Alexey Vorozhtsov[1] and Michele Modena[2]

1 – CERN – TE Department
1211 Geneva 23 – Switzerland, on leave from JINR 141980-Dubna, Russia
2 – CERN – TE Department
1211 Geneva 23 – Switzerland

Two high field quality quadrupole magnets QF1FF and QD0FF are required for the final focus system of the ATF3. In this paper we focus on the design of the QF1FF magnet. The proposed design is a permanent magnet quadrupole (PMQ) with adjustable strength. Alternative solutions such as conventional electromagnetic quadrupole (EMQ) and a hybrid quadrupole (combination of permanent magnet and electromagnet) are also presented and briefly discussed.

## 1 Introduction

Two high field quality quadrupole magnets QF1FF and QD0FF with the differing integrated field gradients are needed for the final focusing system of ATF3- an Accelerator Test Facility at KEK (Japan).

The magnet parameters, such as aperture, integrated magnetic field gradient and required field quality are determined by beam optics considerations. To avoid increasing of the beam spot size at the interaction point (IP), both skew and normal sextupole, octupole, decapole and dodecapole relative field components are limited to a very small values. Table 1 summarizes the requirements for the QF1FF and the QD0FF magnets [1]. In this paper we focus on the design of the QF1FF magnet.

| Magnet Name | QF1FF | QD0FF | Units |
|---|---|---|---|
| Gradient Nominal / Ultra low | 6.772 / 6.791 | 12.45 / 12.46 | T/m |
| Magnetic length | 475 | | mm |
| Nominal Integrated gradient | 3.226 | 5.919 | T |
| Tuning range | ±5 | | % |
| Aperture radius | > 35 | | mm |
| Good Field Region(GFR) radius | 20 | | mm |
| Field quality requirements | | | |
| Harmonic №: | skew $a_n$ | normal $b_n$ | |
| 3 | 0.124 | 0.748 | units@$r_{GFR}$ |
| 4 | 0.344 | 4.12 | units@$r_{GFR}$ |
| 5 | 0.665 | 2.76 | units@$r_{GFR}$ |
| 6 | 1.57 | 9.82 | units@$r_{GFR}$ |

Table 1: Requirements for the ATF3 final focusing quarupoles QF1FF and QD0FF

## 2   Magnet Design Concepts for the QF1FF

As mentioned above skew and normal relative multipole errors are limited to extremely small values. For an ideal (symmetric) case in a quadrupole magnet the first allowed multipole error is $b_6$, which can be controlled by a well designed pole profile and end chamfers. So called "unallowed" multipole errors n=3, 4 and 5, which are not present in an ideal symmetric case, could appear due to fabrication and assembling errors of the magnet yoke.

Therefore, to satisfy the magnet physics requirements the number of yoke segments should be minimized. This will reduce possible the "unallowed" multipole errors resulting from assembly errors.

In the quadrupole magnet, relative multipole errors vary with radius as $r^{n-2}$ according to the formula:

$$b_n \text{ or } a_n \ @r_{GFR} = \left(\frac{B_n \text{ or } A_n}{B_2}\right)_{r_{GFR}} = \left(\frac{B_n \text{ or } A_n}{B_2}\right)_{r_{POLE}} \cdot \left(\frac{r_{GFR}}{r_{POLE}}\right)^{n-2} \quad (1)$$

Thus, larger magnet aperture for the same Good Field Region radius has the advantage of more relaxed mechanical tolerances. However, this approach is not efficient for the sextupole components $a_3$ and $b_3$ due to their linear dependence on radius. Therefore, to suppress the sextupole components an additional trimming of each pole independently is required.

From the magnet design point of view the three options could be investigated:

- Conventional electromagnetic quadrupole (EMQ)
- An adjustable permanent magnet quadrupole (PMQ)
- Hybrid quadrupole (combination of a permanent magnet and an electromagnet)

In this paper as a base line, we are mainly investigated the permanent magnet option, while for the EMQ and the hybrid magnet only a preliminary design based on analytical formulas was performed. For all magnet types the aperture radius was increased from 35 mm to 40 mm due to the reason mentioned above.

## 3   QF1FF Permanent Magnet Quadrupole (PMQ)

The magnet cross-section of the PMQ looks like a classical conventional quadrupole magnet where the coils are replaced by the permanent magnet (p.m.) blocks as flux generators. Each of 4 p.m. blocks generates the flux independently, and field accuracy depends on the uniformity of the easy axis orientation and field strength of the blocks. $Sm_2Co_{17}$ was chosen as a material for the permanent magnet blocks due to its radiation hardness and its weaker temperature dependence.

The proposed design has a limited ability to adjust the field gradient; also it allows reducing the possible sextupole field errors by regularization of the magnetic reluctance on each individual pole, performed by tuning blocks, rectangular shaped, per pole, independently movable. Non-magnetic spacers of different thickness provide a correct position of the tuning blocks.

For the ideal symmetric case the field quality inside the magnet aperture is controlled by soft ferromagnetic pole tips of suitable shape. In addition, these pole tips will smooth the effects of possible p.m. inequalities.

A correct assembly of permanent blocks and pole tips is guaranteed by the aluminum core made of one or two pieces. The pole tips that are glued together with the permanent blocks will be inserted into the aluminum core. This approach enhances the magnet rigidity and simplifies the assembly procedure. Also, this made it possible to reduce the potential assembly errors and the resulting random multipole errors.

The return yoke consists of four pieces made of soft ferromagnetic steel. The return yoke pieces will be mounted, with guiding pins and driving screws on the aluminum core. The main parameters of the PMQ are listed in Table 2 and the mechanical layout of the PMQ is shown in Figure 1 where:

1. P.M. blocks - $Sm_2Co_{17}$
2. Aluminium core
3. Return yoke - Soft ferromagnetic steel.
4. Pole tips - Soft ferromagnetic steel.
5. Tuning blocks - Soft ferromagnetic steel.
6. Non-magnetic spacers- Stainless steel.

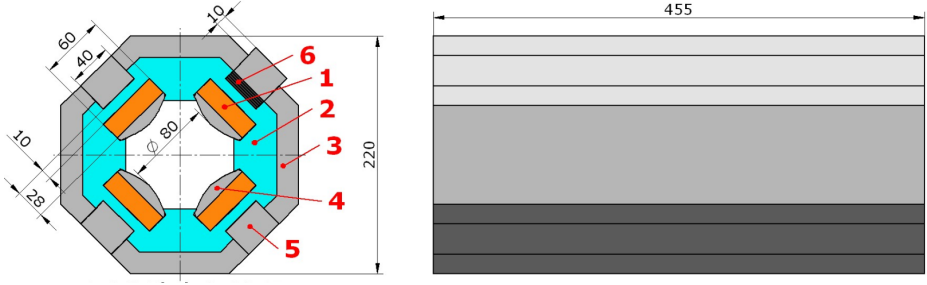

Figure 1: QF1FF PMQ layout

| Aperture radius | 40 mm |
|---|---|
| Magnet height ×width × length | 220 ×220×455 mm |
| Effective length | 474 mm |
| Integrated field gradient | 3.225 T |
| Tuning range | 12.5 % |
| Material of permanent blocks | $Sm_2Co_{17}$ |
| Material of soft ferromagnetic parts | AISI 1010 |
| Allowed (symmetric case) harmonic contents at the GFR radius | < 0.4 units |

Table 2: QF1FF PMQ main parameters

## 3.1 Field Computation for Permanent Magnet Quadrupole (PMQ)

### 3.1.1 Field computation for a symmetric case

Magnetic field calculations of the PMQ were performed with the Opera-2D/ST and Opera-3D/TOSCA programs. The 2D code was used to design the pole profile, return yoke cross-section, permanent block size and position. The magnet end termination was optimized with the 3D code.

Due to symmetry only 1/8 of the magnet geometry was modeled (see, Figure 2). The boundary conditions were chosen in a way that the flux lines were perpendicular to the horizontal middle plane and parallel to the symmetry axis and the limiting edge of the model.

The quadrupole pole tip profile was optimized such that the unwanted higher harmonic field contribution was minimized for the central part of the magnet. The pole tip profile is an arc with the radius of 41.5 mm and angular span of $30^0$ in the middle, two arcs with the radius of 60 mm and angular span of $5.5^0$ at both sides, which are extended smoothly to the tangential lines. These lines act as a shim to increase the gradient at the end of the good field region. The tapered pole ends are rounded with the radius of 3 mm in order to avoid the flux line concentration and the saturation effects in this region. The final pole profile is illustrated in Figure 3

The harmonics content were obtained from the model calculations by Fourier analysis of the radial magnetic field component Br on circle with a radius of 20 mm (good field region boundary), see Figure 4.

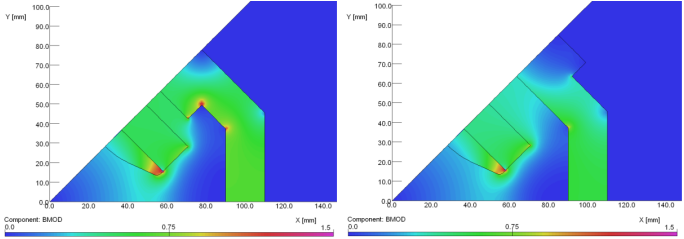

Figure 2: QF1FF PMQ, 2D magnetic field distribution for 2 cases: maximum field gradient (left figure), minimum field gradient (right figure)

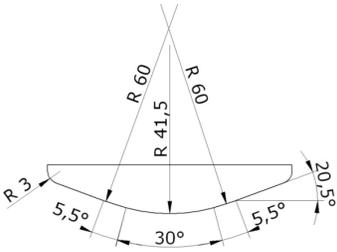

Figure 3: Pole tip profile for the QF1FF PMQ.

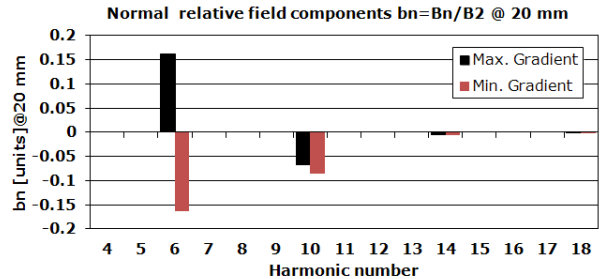

Figure 4: Normal relative field components at GFR radius for 2 simulation cases in 2D

In practice, dynamic of particle beams through the quadrupole magnet is determined by the integrated field gradient, ∫Gdz. Therefore, to estimate the integrated magnetic field characteristics a 3D calculation is required. The calculation was done with the Opera-3D/TOSCA program. Figure 5 shows the OPERA-3D model with the field distribution on the magnet surface. The field gradient distribution along the z-axis is shown in Figure 6. The contributions from the end sections to the higher harmonics were minimized by introducing a $45^0$ chamfer on the pole ends. The values of the integrated harmonics content were obtained from the model calculations by Fourier analysis of the radial field component Br integrated on a cylindrical surface with radius of 20 mm (Good Field Region boundary) and the length L=1000 mm (-500 mm < z < 500 mm). The chamfer size was optimized and the minimal dodecapole field component $b_6$ corresponds to the chamfer height of 8 mm. As shown in Figure 7 the dodecapole component $b_6$ is equal to 0.35 units at GFR radius that is much less than the allowed value. The integrated field gradient error stays below 0.01% inside the good field region for the selected chamfer height see Figure 8. Therefore, the 3D magnetic field calculation confirms the validity of this design in case of an ideal geometry.

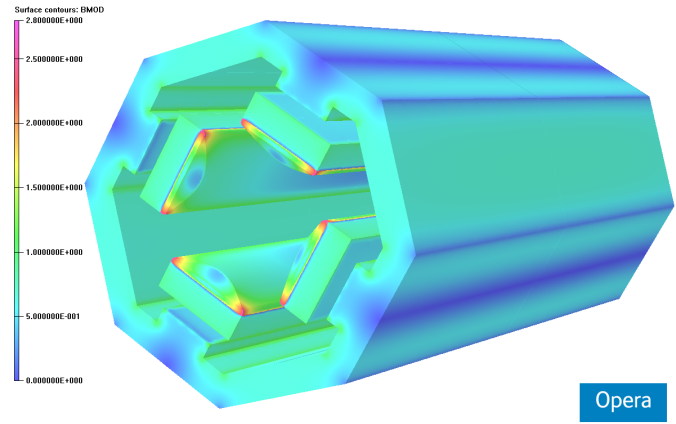

Figure 5: QF1FF PMQ, field distribution on the magnet surface OPERA 3D model.

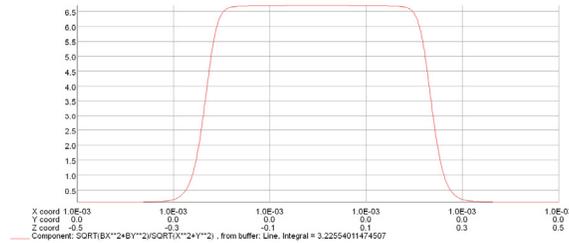

Figure 6: Field gradient distribution along Z-axis

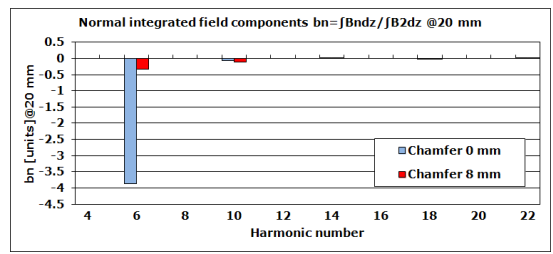

Figure 7: Integrated relative field harmonics for the chamfer height of 0 mm and 8 mm.

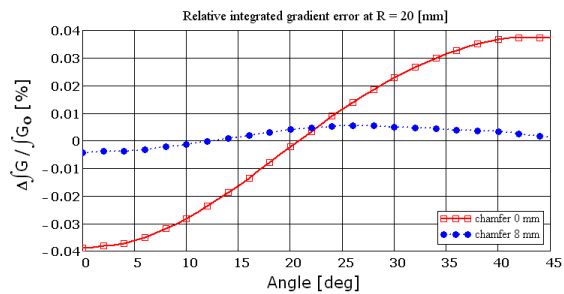

Figure 8: Relative integrated field gradient error [%] at GFR radius for the chamfer height of 0 mm and 8 mm.

### 3.1.2 Easy axis orientations errors

The effects of possible permanent blocks inequalities due to the easy axis orientation errors were computed by introducing an angular deviation from the nominal value in the range of ± $2^0$, representing an upper limit according to the permanent magnet blocks manufacturer (see Figure 9). A full $360^0$ OPERA 2D model was used for this study. The results of the calculations for 2 non-symmetrical cases in comparison to the ideal case are presented in Figure 10. It was found that the field multipole errors stay below the permissible level inside the GFR and the magnet design with the soft ferromagnetic pole tips was not sensitive to such errors.

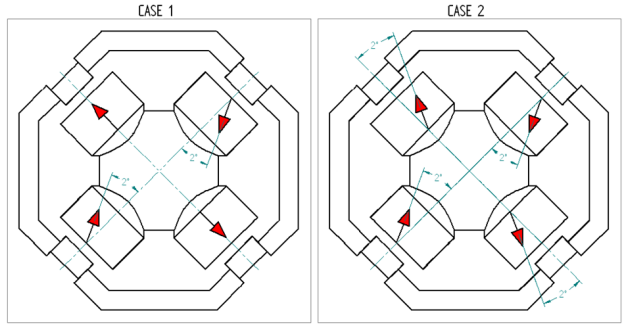

Figure 9: Configuration of easy axis orientation errors

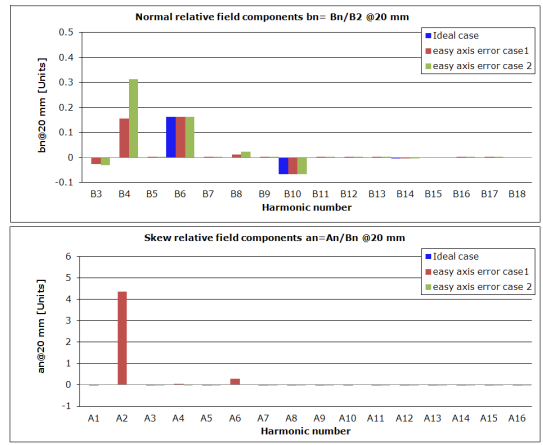

Figure 10: Effect on an error in the magnetization angle

### 3.1.3  *Sextupole components suppression by the tuning blocks*

As mentioned above, to minimize both normal and skew sextupole components an additional trimming is required. For the PMQ this can be done by tuning of magnetic reluctance on each individual pole performed by radial offset of the tuning blocks (at $45^0$, $135^0$, $225^0$, and $315^0$) as shown in Figure 11. According to K. Halbach theory of the first order perturbation in iron dominated magnets, this introduces higher order field components. Formulas and tables given in [2] were used to compute the spectrum of normalized multipoles due to the tuning blocks offset. Figure 12 shows the vectors of dipole and sextupole field components in normal-skew phase plane, generated by tuning blocks for the four cases mentioned above. It is shown that, with help of the tuning blocks we can introduce skew and normal sextupole field components with any phase. Also, the combination of the cases 1 and 4 or 2 and 3 will give only skew sextupole and of the cases 1 and 2 or 3 and 4 only normal sextupole. It should be noted that the dipole field components are also generated in these cases that leads to a shift of the magnetic center. However, if the dipole and the sextupole errors come from the same source, they both will be suppressed by this method. Therefore, the introduction of field components of the same magnitude, but with a phase shift of $180^0$ could correct the possible manufacturing and assembling errors.

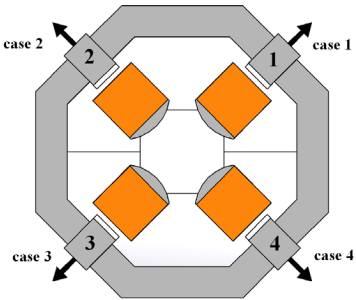

Figure 11: Tuning blocks offset for 4 cases.

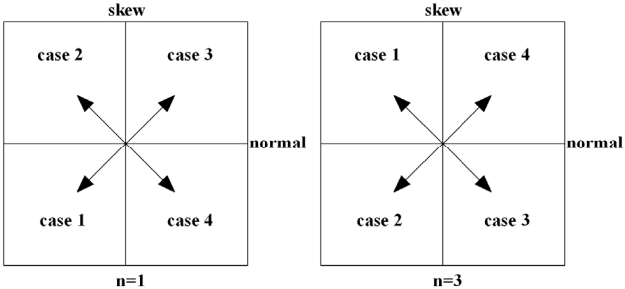

Figure 12: Vectors of field components n=1 and n=3 in normal-skew phase plane generated by tuning blocks offset for 4 cases.

## 4  Alternative solutions

### 4.1  Electromagnetic Quadrupole (EMQ)

The one of the possible alternative solution is an electromagnetic quadrupole magnet. The magnet aperture radius is 40 mm and the overall dimensions of the magnets are 330 mm (width) × 330 mm (height) × 541 mm (length). To achieve a required integrated field gradient, each coil should provide a 4560 A of current-turns product (NI). Each coil consists of 13 turns made of hollow copper conductor with a square cross section of 10 mm × 10 mm and a circular cooling hole with a diameter of 5 mm. The preliminary mechanical layout of the EMQ design is shown in Figure 13. In order to permit an installation of the coils around each pole the magnet core should be divided into four parts by the horizontal and vertical symmetric planes. The main magnet parameters are summarized in the Table 3.

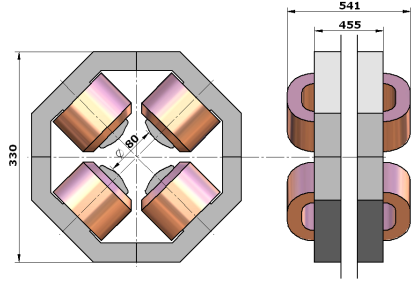

Figure 13: QF1FF EMQ preliminary layout

| Parameters | units | |
|---|---|---|
| Aperture radius | mm | 40 |
| Effective length | mm | 475 |
| Field gradient | T/m | 6.791 |
| Yoke length | mm | 455 |
| Conductor dimensions | mm | $10 \times 10$, $\varnothing=5$ |
| Number of turns per coil | | 13 |
| Ampere turns per pole | A | 4560 |
| Current density | A/mm$^2$ | 4.42 |
| Total resistance | mOhm | 12.3 |
| Voltage | V | 4.3 |
| Power | kW | 1.5 |
| Coolant velocity | m/s | 1.1 |
| Cooling flow per circuit | l/min | 1.28 |
| Pressure drop | bar | 2.3 |
| Temperature rise | K | 17 |

Table 3: QF1FF EMQ main parameters

### 4.2 Hybrid Quadrupole

The design of the hybrid quadrupole magnet is based on the same principles as for the permanent quadrupole magnet (PMQ), but the tuning blocks were replaced by the trim coils. This approach allows tuning the field gradient remotely by changing the excitation current of the trim coils. However to suppress the possible multipole errors four independent power supplies are required.

The preliminary mechanical layout of the PMQ is shown Figure 14 where:
1. P.M. blocks - $Sm_2Co_{17}$
2. Aluminium core
3. Return yoke - Soft ferromagnetic steel.
4. Pole tips - Soft ferromagnetic steel.
5. Pole -Soft ferromagnetic steel.
6. Trim coils.

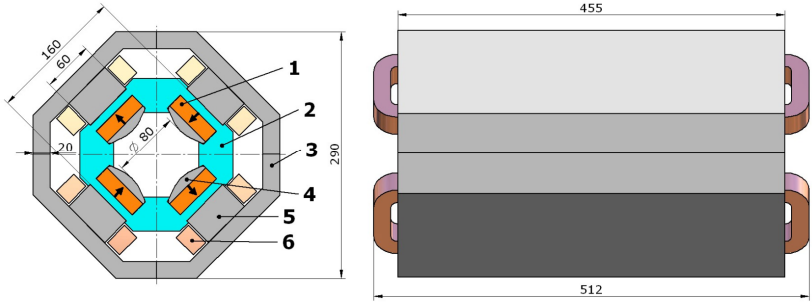

Figure 14: QF1FF Hybrid quadrupole preliminary layout

| Aperture radius | 40 mm |
|---|---|
| Magnet length | 512 mm |
| Yoke height ×width × length | 290 ×290×455 mm |
| Effective length | 474 mm |
| Nom. field gradient | 6.9 T/m |
| Nom. Integrated field gradient | 3.27 T |
| Tuning range | ± 4.6 % |
| Number of turns per pole | 35 |
| Conductor size | 4 mm x 4 mm |
| Current | ± 20 A |
| Max. current density | 1.2 A/mm$^2$ |
| Cooling | Air, natural convention |

Table 4: QF1FF Hybrid quadrupole main parameters

## 5 Summary

Three possible solutions such as an electromagnetic quadrupole magnet (EMQ), an adjustable permanent quadrupole magnet (PMQ) and a hybrid quadrupole magnet were investigated as candidates for a final focus lens in the ATF3.
The PMQ solution looks more preferable over than the EMQ and the hybrid magnet due to the following reasons:

- Compactness of the PMQ structure
- No vibration of the magnet induced by an active water cooling system which is required for EMQ option.
- No failures in the power supplies, which increases the reliability of the magnet.
- Maintenance of coils, cables and power supplies is not required.
- Set to zero operational costs related to electrical energy and cooling systems.
- PMQ can be assembled from one or two pieces, while for the EMQ option only four pieces yoke structure is possible.
- The proposed PMQ design has an ability to suppress the possible higher order multipole errors performed by the tuning blocks, while for the EMQ and the hybrid cases an additional trim coils and four independent power supplies are needed.